\documentstyle[amstex,aps]{revtex}
\input epsf.tex
\begin{document}
\title{The Dyer-Roeder distance in quintessence cosmology and the estimation of $%
H_{0}$ through time-delays.}
\author{Fabio Giovi, Franco Occhionero and Luca Amendola}
\address{Osservatorio Astronomico di Roma, \\
Viale del Parco Mellini 84, \\
00136 Roma, Italy}
\date{\today }
\maketitle

\begin{abstract}
We calculate analytically and numerically the Dyer-Roeder distance in
perfect fluid quintessence models and give an accurate fit to the numerical
solutions for all the values of \ the density parameter and the quintessence
equation of state. Then we apply our solutions to the estimation of $H_{0}$
from multiple image time delays and find that the inclusion of quintessence
modifies sensibly the likelihood distribution of $H_{0}$ , generally
reducing the best estimate with respect to a pure cosmological constant.
Marginalizing over the other parameters ($\Omega _{m}$ and the quintessence
equation of state) , we obtain $H_{0}=71\pm 6$ km/sec/Mpc for an empty beam
and $H_{0}=64\pm 4$ km/sec/Mpc for a filled beam. We also discuss \ the
future prospects for distinguishing quintessence from a cosmological
constant with time delays.
\end{abstract}

\section{Introduction}

Quintessence (Caldwell et al. 1998) or dark energy is a new component of the
cosmic medium that has been introduced in order to explain the dimming of
distant SNIa \ (Riess et al. 1998; Perlmutter et al. 1999) through an
accelerated expansion while at the same time saving the inflationary
prediction of a flat universe. The recent measures of the CMB at high
resolution (Lange et al. 2000, de Bernardis et al. 2000, Balbi et al. 2000)
have added to the motivations for a conspicuous fraction of unclustered dark
energy with negative pressure. In its simplest formulation (see e.g.
Silveira \& Waga 1997), the quintessence component can be modeled as a
perfect fluid with equation of state 
\begin{equation}
p_{q}=\left( \frac{m}{3}-1\right) \rho _{q}  \label{eq stato q}
\end{equation}
with $m$ in the range $0\leq m<3$ ($m<2$ for acceleration). When $m=0$ we
have pure cosmological constant, while for $m=3$ we reduce to the ordinary
pressureless matter. The case $m=2$ mimicks a universe filled with cosmic
strings (see e.g. Vilenkin 1984). More realistic models possess an effective
equation of state that changes with time, and can be modeled by scalar
fields (Ratra \& Peebles 1988, Wetterich 1995, Frieman et al. 1995, Ferreira
and Joyce 1998), possibly with coupling to gravity or matter (Baccigalupi,
Perrotta \& Matarrese 2000, Amendola 2000).

The introduction of the new component modifies the universe expansion and
introduces at least a new parameter, $m$, in cosmology. Most deep
cosmological tests, from large scale structure to CMB, from lensing to deep
counting, are affected in some way by the presence of the new field. Here we
study how a perfect fluid quintessence affects the Dyer-Roeder (DR)
distance, a necessary tool for all lensing studies (Dyer \& Roeder 1972,
1974). The assumption of constant equation of state is at least partially
justified by the relatively narrow range of redshift we are considering, $%
z\in (0.4,2.5)$. We rederive the DR equation in quintessence cosmology, we
solve it analytically whenever possible, and give a very accurate analytical
fit to its numerical solution. Finally, we apply the DR solutions to a
likelihood determination of $H_{0}$ through the observations of time-delays
in multiple images. The dataset we use is composed of only six time-delays,
and does not allow to test directly for quintessence; however, we will show
that inclusion of such cosmologies may have an important impact on the
determination of $H_{0}$ with this method. For instance, we find that $H_{0}$
is smaller that for a pure cosmological constant. In this work we confine
ourselves to flat space and extremal values of the beam parameter $\alpha $;
in a paper in preparation we extend to curved spaces and general $\alpha $.

\section{The Dyer-Roeder distance in quintessence cosmology}

In this section we derive the DR distance in quintessence cosmology, find
its analytical solutions, when possible, and its asymptotic solutions.
Finally, we give a very accurate analytical fit to the general numerical
solutions as a function of $\Omega _{m}$ and $m$.

First of all, let us notice that when quintessence is present, the Friedmann
equation (the $0,0$ component of the Einstein equations in a flat FRW
metric) becomes (in units $G=c=1$)

\begin{equation}
\left( \frac{\dot{a}}{a}\right) ^{2}=H_{0}^{2}\left[ \Omega
_{m}a^{-3}+\left( 1-\Omega _{m}\right) a^{-m}\right] .
\end{equation}
where $H_{0}$ is the present value of the Hubble constant, $\Omega _{m}$ the
present value of the matter density parameter, and where the scale factor is
normalized to unity today. In terms of the redshift $z$ we can write 
\begin{equation}
\frac{\dot{a}}{a}=H_{0}E\left( z\right) .  \label{H(z)}
\end{equation}
where 
\[
E^{2}\left( z\right) =\Omega _{m}\left( 1+z\right) ^{3}+\left( 1-\Omega
_{m}\right) \left( 1+z\right) ^{m} 
\]

The Ricci focalization equation in a conformally flat metric (such as the
FRW metric) with curvature tensor $R_{\alpha \beta }$ is (see e.g.
Schneider, Falco \& Ehlers 1992) 
\begin{equation}
\stackrel{..}{\sqrt{A}}+\frac{1}{2}R_{\alpha \beta }k^{\alpha }k^{\beta }%
\sqrt{A}=0  \label{focricci}
\end{equation}
where $A$ \ is the beam area and $k^{\alpha }=\frac{dx^{\alpha }}{dv}$ is
the tangent vector to the surface of propagation of the light ray, and the
dot means derivation with respect to the affine parameter $v$. Multiplying
the Einstein's gravitational field equation $R_{\alpha \beta }-\frac{1}{2}%
g_{\alpha \beta }R=8\pi T_{\alpha \beta }$ by $k^{\alpha }k^{\beta }$ and
imposing the condition $g_{\alpha \beta }k^{\alpha }k^{\beta }=0$ for the
null geodesic we obtain $R_{\alpha \beta }k^{\alpha }k^{\beta }=8\pi
T_{\alpha \beta }k^{\alpha }k^{\beta }$ ; from (\ref{focricci}) we obtain 
\begin{equation}
\stackrel{..}{\sqrt{A}}+4\pi T_{\alpha \beta }k^{\alpha }k^{\beta }\sqrt{A}%
=0.  \label{focricci2}
\end{equation}
Considering only ordinary pressureless matter and quintessence the
energy-momentum tensor writes 
\begin{equation}
T_{\alpha \beta }=\rho _{m}u_{\alpha }u_{\beta }+\rho _{q}\left( 1+\gamma
\right) u_{\alpha }u_{\beta }-\gamma \rho _{q}g_{\alpha \beta };  \label{Tq}
\end{equation}
multiplying by $k^{\alpha }k^{\beta }$, putting $\gamma =\frac{m}{3}-1$ and
inserting in (\ref{focricci2}) we have 
\begin{equation}
\stackrel{..}{\sqrt{A}}+4\pi \left[ \rho _{m}+\frac{m}{3}\rho _{q}\right] 
\sqrt{A}=0.  \label{focricci3}
\end{equation}
Now, the angular diameter distance $D$ is defined as the ratio between the
diameter of an object and its angular diameter. We have then $D\propto \sqrt{%
A}$. Since $u_{\alpha }k^{\alpha }=\omega =\frac{1}{H_{0}}\left( 1+z\right) $%
, and defining the dimensionless distance $r=DH_{0}$ , Eq. (\ref{focricci3})
writes 
\begin{equation}
\frac{d^{2}r}{d\sigma ^{2}}+\frac{4\pi }{H_{0}^{2}}\left( 1+z\right) ^{2}%
\left[ \rho _{m}+\frac{m}{3}\rho _{q}\right] r=0  \label{focricciaffine}
\end{equation}
where we introduced the affine parameter $\sigma $ defined implicitely by
the relation 
\begin{equation}
g\left( z\right) =\frac{dz}{d\sigma }=\left( 1+z\right) ^{2}E\left( z\right) 
\label{g(z)}
\end{equation}
where $E\left( z\right) $ is defined in Eq. (\ref{H(z)}). Finally we get the
DR equation with the redshift as independent variable 
\begin{equation}
\begin{array}{c}
\left( 1+z\right) ^{2}\left[ \Omega _{m}\left( 1+z\right) ^{3}+\left(
1-\Omega _{m}\right) \left( 1+z\right) ^{m}\right] \frac{d^{2}r}{dz^{2}}+ \\ 
\\ 
+\left( 1+z\right) \left[ \frac{7}{2}\Omega _{m}\left( 1+z\right) ^{3}+\frac{%
m+4}{2}\left( 1-\Omega _{m}\right) \left( 1+z\right) ^{m}\right] \frac{dr}{dz%
}+ \\ 
\\ 
+\left[ \frac{3}{2}\alpha \Omega _{m}\left( 1+z\right) ^{3}+\frac{m}{2}%
\left( 1-\Omega _{m}\right) \left( 1+z\right) ^{m}\right] r=0
\end{array}
\label{drg}
\end{equation}
The constant $\alpha $ in Eq. (\ref{drg}) is the fraction of matter
homogeneously distributed inside the beam: when $\alpha =0$ all the matter
is clustered (empty beam), while for $\alpha =1$ the matter is spread
homogeneously and we recover the usual angular diameter distance (filled
beam). Notice that in our case the ''empty beam'' is actually filled
uniformly with quintessence. Since the actual value of $\alpha $ is unknown
(see however Barber et al. 2000, who argue in favor of $\alpha $ near
unity), we will adopt in the following the two extremal values $\alpha =0$
and $\alpha =1$. The appropriate boundary conditions are (see e.g.
Schneider, Falco \& Ehlers 1992) 
\begin{equation}
\left\{ 
\begin{array}{l}
r\left( z_{1},z_{1}\right) =0 \\ 
\\ 
\frac{dr\left( z_{1},z\right) }{dz}|_{z_{1}}=\left( 1+z_{1}\right) ^{-1}%
\left[ \Omega _{m}\left( 1+z_{1}\right) ^{3}+\left( 1-\Omega _{m}\right)
\left( 1+z_{1}\right) ^{m}\right] ^{-1/2}
\end{array}
\right.   \label{condizioni al contorno z}
\end{equation}
Defining $r\left( 0,z\right) =r\left( z\right) $ these become 
\begin{equation}
\left\{ 
\begin{array}{c}
r\left( 0\right) =0 \\ 
\\ 
\frac{dr}{dz}|_{0}=1
\end{array}
\right.   \label{condizioni al contorno}
\end{equation}

{\bf Analytical solutions}. Eq. (\ref{drg}) has an analytical solution only
for some values of $\alpha $ and $m$. Here we list all the known analytical
solutions. All the cases for $\alpha =0$ and $m\neq 0$ or $\Omega _{m}\neq
0,1$ are new solutions. For the case $m=0$ see also Demiansky et al. (2000).

Case \underline{$\alpha =0${\bf , }$m=0$}\newline
When $m=0$ the quintessence fluid is the cosmological constant; the DR
equation becomes 
\begin{equation}
\begin{array}{c}
\left( 1+z\right) \left[ \Omega _{m}\left( 1+z\right) ^{3}+\left( 1-\Omega
_{m}\right) \right] \frac{d^{2}r}{dz^{2}}+ \\ 
+\left[ \frac{7}{2}\Omega _{m}\left( 1+z\right) ^{3}+2\left( 1-\Omega
_{m}\right) \right] \frac{dr}{dz}=0
\end{array}
\end{equation}
whose solution in integral form is 
\begin{equation}
r\left( z\right) =\int\limits_{0}^{z}\frac{dw}{\left( 1+w\right) ^{2}\sqrt{%
\Omega _{m}\left( 1+w\right) ^{3}+\left( 1-\Omega _{m}\right) }}
\label{a0m0}
\end{equation}
that is 
\begin{equation}
\begin{tabular}{c}
$r\left( z\right) =\frac{1}{\sqrt{1-\Omega _{m}}}\left( F\left[ -\frac{1}{3},%
\frac{1}{2};\frac{2}{3},\frac{\Omega _{m}}{\Omega _{m}-1}\right] -\frac{1}{%
1+z}F\left[ -\frac{1}{3},\frac{1}{2};\frac{2}{3},\frac{\Omega _{m}}{\Omega
_{m}-1}(1+z)^{3}\right] \right) $%
\end{tabular}
\end{equation}
where $F\left[ a,b;c,x\right] $ is the Gauss hypergeometric function.

Case \underline{$\alpha =0${\bf , }$m=1$}\newline
The solution is 
\begin{equation}
\begin{tabular}{c}
$r\left( z\right) =\frac{2}{1+z}\frac{\sqrt{1-\Omega _{m}+\Omega _{m}\left(
1+z\right) ^{2}}F\left[ -\frac{1}{4},1;\frac{5}{4},\frac{\Omega _{m}}{\Omega
_{m}-1}\right] -\sqrt{1+z}F\left[ -\frac{1}{4},1;\frac{5}{4},\frac{\Omega
_{m}}{\Omega _{m}-1}\left( 1+z\right) ^{2}\right] }{\left( 2\Omega
_{m}-1\right) F\left[ -\frac{1}{4},1;\frac{5}{4},\frac{\Omega _{m}}{\Omega
_{m}-1}\right] +\frac{4}{5}\frac{\Omega _{m}}{\Omega _{m}-1}F\left[ \frac{3}{%
4},2;\frac{9}{4},\frac{\Omega _{m}}{\Omega _{m}-1}\right] }$%
\end{tabular}
\end{equation}
\newline
\underline{Case $\alpha =0${\bf , }$m=2$}\newline
The solution can be written in terms of the Meijer $G$ function, but the
expression is so complicated that it is not worth reporting it here. \newline
Case \underline{$\alpha =0${\bf , }$m=3$}\newline
Now quintessence coincides with ordinary matter; however, the choice $\alpha
=0$ implies that the ordinary matter is completely clustered, while
quintessence remains homogeneous. The solution is 
\begin{equation}
r\left( z\right) =\frac{2\left( 1+z\right) ^{-\frac{5}{4}}}{S\left( \Omega
_{m}\right) }\left[ \left( 1+z\right) ^{\frac{1}{4}S\left( \Omega
_{m}\right) }-\left( 1+z\right) ^{-\frac{1}{4}S\left( \Omega _{m}\right) }%
\right]  \label{soluzione alfa0 m3}
\end{equation}
where 
\begin{equation}
S\left( \Omega _{m}\right) =\sqrt{\frac{1+23\Omega _{m}-24\Omega _{m}^{2}}{%
1-\Omega _{m}}}
\end{equation}
Cases \underline{$\alpha =0${\bf , }$\Omega _{m}=0,1$}\newline
For $\Omega _{m}=0$ the DR equation is 
\begin{equation}
\left( 1+z\right) ^{2}\frac{d^{2}r}{dz^{2}}+\frac{m+4}{2}\left( 1+z\right) 
\frac{dr}{dz}+\frac{m}{2}r=0
\end{equation}
and its solution is 
\begin{equation}
r\left( z\right) =\left\{ 
\begin{tabular}{cc}
$\frac{2}{m-2}\left[ \frac{1}{1+z}-\frac{1}{\left( 1+z\right) ^{m/2}}\right] 
$ & $m\neq 2$ \\ 
&  \\ 
$\frac{1}{1+z}\ln \left( 1+z\right) $ & $m=2$%
\end{tabular}
\right.
\end{equation}
while for $\Omega _{m}=1$ we get 
\begin{equation}
\left( 1+z\right) \frac{d^{2}r}{dz^{2}}+\frac{7}{2}\frac{dr}{dz}=0
\end{equation}
and 
\begin{equation}
r\left( z\right) =\frac{2}{5}\left( 1-\frac{1}{\left( 1+z\right) ^{5/2}}%
\right)
\end{equation}
The case $\alpha =1$ is actually the standard angular diameter distance in a
homogeneous universe. We list some particular solution here for completeness
(see also Bloomfield-Torres \& Waga 1996, Waga \& Miceli 1999).

Case \underline{$\alpha =1${\bf , }$m\neq 3$}\newline
The general solution is 
\begin{equation}
r\left( z\right) =\frac{1}{1+z}\int\limits_{0}^{z}\frac{dw}{\sqrt{\Omega
_{m}\left( 1+w\right) ^{3}+\left( 1-\Omega _{m}\right) \left( 1+w\right) ^{m}%
}}  \label{angdiam}
\end{equation}
which gives for any $m$ 
\begin{equation}
\begin{tabular}{c}
$r\left( z\right) =\frac{2}{\left( m-2\right) \sqrt{1-\Omega _{m}}}\left( 
\frac{1}{1+z}F\left[ \frac{m-2}{2\left( m-3\right) },\frac{1}{2};\frac{3m-8}{%
2\left( m-3\right) },\frac{\Omega _{m}}{\Omega _{m}-1}\right] \right. +$ \\ 
$\left. -\frac{1}{\sqrt{\left( 1+z\right) ^{m}}}F\left[ \frac{m-2}{2\left(
m-3\right) },\frac{1}{2};\frac{3m-8}{2\left( m-3\right) },\frac{\Omega _{m}}{%
\Omega _{m}-1}\left( 1+z\right) ^{3}\right] \right) $%
\end{tabular}
\end{equation}

Case \underline{$\alpha =1${\bf , }$m=3$}\newline
The solution reduces to 
\begin{equation}
r\left( z\right) =2\frac{\sqrt{1+z}-1}{\sqrt{\left( 1+z\right) ^{3}}}
\end{equation}
Case \underline{$\alpha =1${\bf , }$\Omega _{m}=0;1$}\newline
For $\Omega _{m}=0$ we get 
\begin{equation}
r\left( z\right) =\frac{1}{1+z}\int\limits_{0}^{z}\frac{dw}{\sqrt{\left(
1+w\right) ^{m}}}
\end{equation}
(identical to the case $\alpha =0${\bf , }$\Omega _{m}=0$). For $\Omega
_{m}=1$\ we reduce to the case $\alpha =1${\bf , }$m=3$.

{\bf Asymptotic limits.} The DR equation can be solved analytically in the
limit of small or large $z$. Here we give these limits; they will be used to
produce accurate analytical fits to the numerical solutions. For small $z$
and $\alpha =0$ we have 
\begin{equation}
\frac{d^{2}r_{s}}{dz^{2}}+\left[ \frac{7}{2}\Omega _{m}+\frac{m+4}{2}\left(
1-\Omega _{m}\right) \right] \frac{dr_{s}}{dz}+\frac{m}{2}\left( 1-\Omega
_{m}\right) r_{s}=0  \label{drgpiccoli}
\end{equation}
whose solution is 
\begin{equation}
r_{s}\left( z\right) =\frac{\sinh \left[ A\left( \Omega _{m},m\right) z%
\right] }{A\left( \Omega _{m},m\right) }\exp \left\{ -\left[ 3\Omega
_{m}+m\left( 1-\Omega _{m}\right) +4\right] \frac{z}{4}\right\}
\end{equation}
where 
\begin{equation}
A\left( \Omega _{m},m\right) =\frac{\sqrt{16+\left[ 3\Omega _{m}+m\left(
1-\Omega _{m}\right) \right] ^{2}+24\Omega _{m}}}{4}
\end{equation}
or, to second order in $z$ 
\begin{equation}
r_{s}\left( z\right) \approx z-\frac{1}{4}\left[ 4+m\left( 1-\Omega
_{m}\right) +3\Omega _{m}\right] z^{2}+O\left( z^{3}\right)  \label{second}
\end{equation}
For completeness, we quote also the result in the case $\alpha \neq 0$: 
\begin{equation}
r_{s}^{\alpha }\left( z\right) =\frac{\sinh \left[ A^{\alpha }\left( \Omega
_{m},m,\alpha \right) z\right] }{A^{\alpha }\left( \Omega _{m},m,\alpha
\right) }\exp \left\{ -\left[ 3\Omega _{m}+m\left( 1-\Omega _{m}\right) +4%
\right] \frac{z}{4}\right\}
\end{equation}
where 
\begin{equation}
A^{\alpha }\left( \Omega _{m},m,\alpha \right) =\frac{\sqrt{16+\left[
3\Omega _{m}+m\left( 1-\Omega _{m}\right) \right] ^{2}+24\left( 1-\alpha
\right) \Omega _{m}}}{4}.
\end{equation}
The limit to second order is identical to Eq. (\ref{second}), which shows
that for small redshift the degree of emptiness is not relevant.

For large $z$ (and $m\neq 3,\Omega _{m}\neq 0,1$) the DR equation for $%
\alpha =0$ reduces to 
\begin{equation}
\left( 1+z\right) ^{2}\frac{d^{2}r_{l}}{dz^{2}}+\frac{7}{2}\left( 1+z\right) 
\frac{dr_{l}}{dz}+\frac{m}{2}\frac{1-\Omega _{m}}{\Omega _{m}}r_{l}=0
\label{drggrandi}
\end{equation}
If we define $r_{l}\left( z\right) =\frac{G\left( z\right) }{\left(
1+z\right) ^{5/4}}$ we obtain the equation 
\begin{equation}
\left( 1+z\right) ^{2}\frac{d^{2}G}{dz^{2}}+\left( 1+z\right) \frac{dG}{dz}+%
\left[ \frac{m}{2}\frac{1-\Omega _{m}}{\Omega _{m}}\left( 1+z\right) ^{m-3}-%
\frac{25}{16}\right] G=0
\end{equation}
whose solution can be written in terms of Bessel functions of first kind: 
\begin{equation}
r_{l}\left( z\right) =\left\{ 
\begin{tabular}{cc}
$\frac{1}{\left( 1+z\right) ^{5/4}}\left[ C_{1}J_{-\nu }\left( y\right)
+C_{2}\Gamma (1+\nu )\left( \frac{m}{2(m-3)^{2}}\frac{1-\Omega _{m}}{\Omega
_{m}}\right) ^{-\nu /2}J_{\nu }\left( y\right) \right] $ & $m\neq 0$ \\ 
&  \\ 
$C_{2}-\frac{2}{5\left( 1+z\right) ^{5/2}}C_{1}$ & $m=0$%
\end{tabular}
\right.  \label{rlarge}
\end{equation}
where 
\begin{gather*}
\nu =\frac{5}{2\left( m-3\right) } \\
\\
y=\sqrt{\frac{2m}{\left( m-3\right) ^{2}}\frac{1-\Omega _{m}}{\Omega _{m}}}%
\left( 1+z\right) ^{\frac{m-3}{2}}
\end{gather*}

\bigskip Notice that the large-$z$ limit is always a constant: 
\[
\lim_{z\rightarrow \infty }r_{l}(z)=C_{2} 
\]

Since in the limit of large $z$ the $C_{1}$ term in (\ref{rlarge}) is
negligible, we will consider in the following only the $C_{2}$ term. Again
for completeness we quote the analogous result for any $\alpha \neq 0$: 
\begin{equation}
r_{l}^{\alpha }\left( z\right) =\frac{1}{\left( 1+z\right) ^{5/4}}\left[
C_{1}\left( 1+z\right) ^{-\frac{\sqrt{25-24\alpha }}{4}}+C_{2}\left(
1+z\right) ^{\frac{\sqrt{25-24\alpha }}{4}}\right]
\end{equation}

{\bf Numerical fits.} The rest of this section focuses on the $\alpha =0$
case, for which we do not have a closed solution. Although in the
application of the next section we employ the exact numerical solutions, it
may prove practical to produce analytical fits. We use the asymptotic limits
above to build an analytical fit to the full numerical solution, valid for
all $\Omega _{m}$ and all $m$. We search a fit in the form 
\begin{equation}
r_{fit}\left( z\right) =r_{s}\left( z\right) T_{s}\left( z\right)
+r_{l}\left( z\right) T_{l}\left( z\right)
\end{equation}
where the function $T_{s}(z)$ is a step-like curve chosen so as to
interpolate from $1$ to $0$ as $z$ goes from $0$ to infinity, and $T_{l}(z)$
interpolates similarly from $0$ to $1.$ We chose 
\begin{equation}
\begin{tabular}{c}
$T_{s}\left( z\right) =\frac{1}{2}\left[ 1-\tanh \left( \frac{1}{\Delta }\ln
\left( \frac{z}{z_{0}}\right) \right) \right] $ \\ 
\\ 
$T_{l}\left( z\right) =\frac{1}{2}\left[ 1+\tanh \left( \frac{1}{\Delta }\ln
\left( \frac{z}{z_{0}}\right) \right) \right] $%
\end{tabular}
\end{equation}
so that the transition occurs at $z_{0}$ and $\Delta $ sets its steepness.
We have then three parameters $\left( C_{2},z_{0},\Delta \right) $ to fit as
functions of $\Omega _{m}$ and $m$. The result for $C_{2}$ in terms of a
simple polynomial fit is 
\begin{gather}
C_{2}=-0.034\Omega _{m}^{2}m^{2}-0.304\Omega _{m}^{2}m+0.112\Omega
_{m}m^{2}+0.339\Omega _{m}^{2} \\
-0.078m^{2}+0.397\Omega _{m}m-0.6\Omega _{m}-0.117m+0.682.  \nonumber
\end{gather}
For $z_{0}$ and $\Delta $ we find convenient to use the following functional
form: 
\begin{equation}
\left( a_{5}m^{5}+a_{4}m^{4}+a_{3}m^{3}+a_{2}m^{2}+a_{1}m+a_{0}\right) \sinh
\left( b_{2}\Omega _{m}^{2}+b_{1}\Omega _{m}+b_{0}\right)
\end{equation}
The values of the fit parameters for $z_{0}$ and $\Delta $ are listed in
Table I. 
\begin{gather*}
\text{Table I} \\
\begin{tabular}{|c|c|c|c|c|c|c|c|c|c|}
\hline
& $a_{5}$ & $a_{4}$ & $a_{3}$ & $a_{2}$ & $a_{1}$ & $a_{0}$ & $b_{2}$ & $%
b_{1}$ & $b_{0}$ \\ \hline
$z_{0}$ & {\small -0.340} & {\small 1.775} & {\small -3.324} & {\small 2.692}
& {\small -0.884} & {\small 0.132} & {\small 2.005} & {\small -3.025} & 
{\small 4.935} \\ \hline
$\Delta $ & {\small -0.280} & {\small 1.687} & {\small -3.747} & {\small %
3.725} & {\small -1.564} & {\small 0.367} & {\small -0.521} & {\small 0.756}
& {\small 1.998} \\ \hline
\end{tabular}
\end{gather*}
Such fits are accurate to better than 5\% over the range $m\in (0,2)$ and $%
\Omega _{m}\in (0,1)$ as can be seen from Fig. 2, where we compare our fits
to a sample of exact numerical solutions.

As a cautionary remark, let us notice that the assumption of a constant $%
\alpha $ over a very large range in redshift is certainly problematic. The
results of the next section, however, are obtained in a relatively narrow
range of redshifts, so that the approximation should be acceptable.

\section{Measuring H$_{0}$ in quintessence cosmology through time-delays}

Now that we are in possess of the general angular diameter distance in
quintessence cosmology we can apply it to real data. As a first application,
in this section we use six observed time-delays to measure $H_{0}$ taking
into account the presence of quintessence fields.

Let us first present the data. There are only seven lens systems with
measured time-delays: B0218+357 (Biggs et al. 1999; Leh\'{a}r et al. 1999;
Patnaik Porcas \& Browne 1995), Q0957+561 (Bar-Kana et al. 1999; Kundi\'{c}
et al. 1997), HE1104-1805 (Courbin, Lidman \& Magain 1998; Wisotzki,
Wucknitz, Lopez \&\ S\o rensen 1998), PG1115+080 (Bar-Kana 1997; Impey et
al. 1998; Schechter et al. 1996), B1600+434 (Burud et al. 2000; Koopmans, de
Bruyn, Xanthopoulos \& Fassnacht 2000), B1608+656 (Fassnacht C. D. et al.
1999; Koopmans \& Fassnacht 1999) and PKS1830-211 (Leh\'{a}r et al. 1999;
Lovell et al. 1998; Wiklind \& Combes 1999). Due to the image multiplicity,
we have in total ten time-delays. The lens model we use, the isothermal lens
of Mao, Witt and Keeton (Mao, Witt \& Keeton 2000) cannot be adapted to
Q0957+561 and B1608+656 so we are left with five lens systems and six
time-delays, as detailed in Table II.

\begin{gather*}
\text{Table II} \\
\begin{tabular}{|c|c|c|c|c|c|c|c|}
\hline
Lens/components & $z_{d}$ & $z_{s}$ & $\Delta t_{ij}$ & $\vartheta
_{i}\left( "\right) $ & $\vartheta _{j}\left( "\right) $ & $f\left( \frac{%
\text{Mpc}}{\text{km/s}}\right) $ & $\frac{\Delta f}{f}$ \\ \hline
\multicolumn{1}{|l|}{B0218+357/BA} & 0.68 & 0.96 & (10.5$\pm $0.2)d & 0.24$%
\pm $0.06 & 0.10$\pm $0.06 & 0.031 & 87\% \\ \hline
\multicolumn{1}{|l|}{HE1104-1805/AB} & 0.77 & 2.32 & (0.7$\pm $0.1)yr & 2.095%
$\pm $0.008 & 1.099$\pm $0.004 & 0.0108 & 16\% \\ \hline
\multicolumn{1}{|l|}{PG1115+080/AB} & 0.31 & 1.72 & (11.7$\pm $1.2)d & 1.147$%
\pm $0.025 & 0.950$\pm $0.004 & 0.0051 & 25\% \\ \hline
\multicolumn{1}{|l|}{PG1115+080/CB} & 0.31 & 1.72 & (25.0$\pm $1.6)d & 1.397$%
\pm $0.004 & 0.950$\pm $0.004 & 0.00433 & 8\% \\ \hline
\multicolumn{1}{|l|}{B1600+434/BA} & 0.42 & 1.59 & (47$\pm $6)d & 1.14$\pm $%
0.05 & 0.25$\pm $0.05 & 0.0064 & 23\% \\ \hline
\multicolumn{1}{|l|}{PKS1830-211/BA} & 0.89 & 2.51 & (26$\pm $5)d & 0.67$\pm 
$0.08 & 0.32$\pm $0.08 & 0.0095 & 64\% \\ \hline
\end{tabular}
\end{gather*}

Adopting the isothermal model, the relation between the distances of the
deflector (subscript $d$) and of the source ($s$) and the time-delay between
two images labelled 1 and 2 in terms of observable quantities is 
\begin{equation}
\Delta t=\frac{1+z_{d}}{2}\frac{D_{d}D_{s}}{D_{ds}}\Delta \vartheta ^{2},
\end{equation}
where 
\begin{equation}
\Delta \vartheta ^{2}=[\vartheta _{1}^{2}-\vartheta _{2}^{2}],
\end{equation}
$\vartheta _{1,2}$ is the angular distance between one of the two images and
the center of the deflector and where

\begin{equation}
D_{a,b}=\frac{1}{H_{0}}r(z_{a},z_{b};\Omega _{m},m).
\end{equation}
To estimate $H_{0}$ we build the likelihood function 
\begin{equation}
-2\log L(H_{0},\Omega _{m},m)=\sum \left[ -\frac{(f_{i,t}(H_{0},\Omega
_{m},m)-f_{i,o})^{2}}{2\sigma _{i}^{2}}\right] ,
\end{equation}
where we separated between quantities that contain theoretical parameters
and purely observational quantities by definining the variables 
\begin{eqnarray}
f_{i,t} &=&\left( \frac{D_{d}D_{s}}{D_{ds}}\right) _{i}, \\
f_{i,o} &=&\left( \frac{2}{1+z_{d}}\frac{\Delta t}{\Delta \vartheta ^{2}}%
\right) _{i},
\end{eqnarray}
and where $\sigma _{i}$ is the total error on $f_{i,o}$, obtained by
standard error propagation (this error is dominated by the uncertainty on
the angular positions $\vartheta _{1,2}$). Then we marginalize over $\Omega
_{m},m$ (defined in the range $0,1$ and $0,3,$ respectively) and produce the
marginalized and normalized likelihood 
\begin{equation}
L(H_{0})=N\int_{0}^{1}d\Omega _{m}\int_{0}^{3}dmL(H_{0},\Omega _{m},m),
\end{equation}
that represents the likelihood of $H_{0}$ (normalized to unity by the
constant $N$) given any possible perfect fluid quintessence model. We also
evaluated $L(m)$ and $L(\Omega _{m})$, marginalizing over the other
variables, but the likelihoods we obtain are too flat to derive any
interesting conclusion on the quintessence parameters. We also estimated the
effect of imposing a Gaussian prior on $\Omega _{m},$ with mean 0.3 and
standard deviation 0.1$.$ We compared the marginalized likelihood with the
likelihood in the ``standard'' cases $\Omega _{m}=0.3$ and $m=0$ (pure
cosmological constant), and $\Omega _{m}=1$ and $m=3$ (no quintessence).
Table III summarizes the likelihood results: here 1,2,3$\sigma $ stand for a
probability of 68,95,99\%, respectively; the first four cases are
marginalized over $\Omega _{m}$ and $m$, and the prior is the above
mentioned Gaussian on $\Omega _{m}$.

The strongest effects are obtained in the empty beam case, $\alpha =0$,
because in this case the quintessence term is dominating in the last term of
the DR equation. The likelihood is shifted to lower values, in comparison to
the two ``standard'' cases (see Fig. 5), with an increase in the variance.
For $\alpha =1$ the shift is less evident and there is a degeneracy between
marginalized likelihood and no quintessence likelihood. Thus the dependency
of $H_{0}$ for $\Omega _{m}$\ and $m$ is strongest in the empty beam case.
With no prior we obtain 
\begin{eqnarray}
H_{0} &=&71\pm 6\text{ km/sec/Mpc if }\alpha =0  \nonumber \\
H_{0} &=&64\pm 4\text{ km/sec/Mpc if }\alpha =1
\end{eqnarray}

\begin{gather*}
\text{Table III} \\
\begin{tabular}{|l|l|l|l|l|}
\hline
Model & Max & $1\sigma $ & $2\sigma $ & $3\sigma $ \\ \hline
$\alpha =0,$margin., no prior & \multicolumn{1}{|c|}{$71$} & 
\multicolumn{1}{|c|}{$65<H_{0}<77$} & \multicolumn{1}{|c|}{$60<H_{0}<83$} & 
\multicolumn{1}{|c|}{$57<H_{0}<88$} \\ \hline
$\alpha =0,$margin., prior & \multicolumn{1}{|c|}{$69$} & 
\multicolumn{1}{|c|}{$64<H_{0}<74$} & \multicolumn{1}{|c|}{$60<H_{0}<80$} & 
\multicolumn{1}{|c|}{$57<H_{0}<85$} \\ \hline
$\alpha =1,$margin., no prior & \multicolumn{1}{|c|}{$64$} & 
\multicolumn{1}{|c|}{$60<H_{0}<69$} & \multicolumn{1}{|c|}{$56<H_{0}<74$} & 
\multicolumn{1}{|c|}{$54<H_{0}<78$} \\ \hline
$\alpha =1$, margin., prior & \multicolumn{1}{|c|}{$64$} & 
\multicolumn{1}{|c|}{$60<H_{0}<69$} & \multicolumn{1}{|c|}{$57<H_{0}<74$} & 
\multicolumn{1}{|c|}{$54<H_{0}<78$} \\ \hline
$\alpha =0$, $\Omega _{m}=0.3$, $m=0$ & \multicolumn{1}{|c|}{$73$} & 
\multicolumn{1}{|c|}{$69<H_{0}<78$} & \multicolumn{1}{|c|}{$65<H_{0}<84$} & 
\multicolumn{1}{|c|}{$62<H_{0}<89$} \\ \hline
$\alpha =0$, $\Omega _{m}=1$, $m=3$ & \multicolumn{1}{|c|}{$75$} & 
\multicolumn{1}{|c|}{$70<H_{0}<80$} & \multicolumn{1}{|c|}{$66<H_{0}<86$} & 
\multicolumn{1}{|c|}{$64<H_{0}<91$} \\ \hline
$\alpha =1$, $\Omega _{m}=0.3$, $m=0$ & \multicolumn{1}{|c|}{$67$} & 
\multicolumn{1}{|c|}{$62<H_{0}<72$} & \multicolumn{1}{|c|}{$59<H_{0}<77$} & 
\multicolumn{1}{|c|}{$57<H_{0}<81$} \\ \hline
$\alpha =1$, $\Omega _{m}=1$, $m=3$ & \multicolumn{1}{|c|}{$63$} & 
\multicolumn{1}{|c|}{$59<H_{0}<68$} & \multicolumn{1}{|c|}{$56<H_{0}<73$} & 
\multicolumn{1}{|c|}{$54<H_{0}<77$} \\ \hline
\end{tabular}
\end{gather*}

Already with six time-delays, the effect of a quintessence cosmology on the
estimation of $H_{0}$ is therefore not negligible. A qualitative idea of how
the method can perform in the future can be gained assuming that the same
six time delays can be estimated with only a 10\% error on the variable $f$.
In this case, we would get not only a better estimation of $H_{0}$ but also
a substantial removal of the degeneracy with respect to $\Omega _{m}$ and $m$%
. This exercise illustrates the potentiality of the method towards a
detection of quintessence and a distinction from a pure cosmological
constant. The results of the simulation for $\alpha =0$ are summarized in
Table IV and in Figure 6. 
\begin{gather*}
\text{Table IV} \\
\begin{tabular}{|l|c|l|l|l|}
\hline
& Max & $1\sigma $ & $2\sigma $ & $3\sigma $ \\ \hline
$H_{0}\left( \text{km/s/Mpc}\right) $ & $62$ & \multicolumn{1}{|c|}{$%
59<H_{0}<66$} & \multicolumn{1}{|c|}{$56<H_{0}<71$} & \multicolumn{1}{|c|}{$%
55<H_{0}<76$} \\ \hline
$\Omega _{m}$ & $0$ & \multicolumn{1}{|c|}{$\Omega _{m}<0.13$} & 
\multicolumn{1}{|c|}{$\Omega _{m}<0.37$} & \multicolumn{1}{|c|}{$\Omega
_{m}<0.59$} \\ \hline
$m$ & $3$ & \multicolumn{1}{|c|}{$m>1.62$} & \multicolumn{1}{|c|}{$m>0.44$}
& \multicolumn{1}{|c|}{$m>0.10$} \\ \hline
\end{tabular}
\end{gather*}

\section{Conclusions}

\bigskip

If 70\% or so of the total matter content is filled by a new component with
negative pressure and weak clustering, all the classical deep cosmological
probes are affected in some way. Here we addressed the question of how the
Dyer-Roeder distance changes when this new component, quintessence, is taken
into account. We have shown that, particularly in the case of empty beam,
the effect of the quintessence is to move the estimate of $H_{0}$ to lower
values with respect to two standard models, and to increase the spread of
the likelihood distribution. This is the first time, to our knowledge, that
a full likelihood analysis of the almost entire set of time-delays available
is performed. As a byproduct of our analysis, we produced fits for a large
range of values of $\Omega _{m}$ and $m$ accurate to within 5\%$.$

The future prospects seem interesting for the time-delay method: with a not
unrealistic increase in accuracy (or in the number of time delays), the
quintessence could be detected and distinguished from a pure cosmological
constant, thanks to the deepness to which lensing effects are observable.

An obvious improvement of our analysis, currently underway, is to
investigate the dependence on $\alpha \neq 0,1$ and on curvature, producing
a fully marginalized likelihood for $H_{0}$.

\section{References}

Amendola L., 2000, Phys. Rev. D62 , 043511, preprint astro-ph/9908440\newline
Baccigalupi C., Perrotta F. \& Matarrese S., 2000, Phys. Rev. D61, 023507,
preprint astro-ph/9906066\newline
Balbi A., et al., 2000, astro-ph/0005124\newline
Barber A., et al., 2000, astro-ph/0002437\newline
Bar-Kana R., 1997, ApJ, 489, 21, preprint astro-ph/9701068\newline
Bar-Kana R., et al., 1999, ApJ, 523, 54\newline
Biggs A. D., et al., 1999, MNRAS, 304 349, preprint astro-ph/9811282\newline
Bloomfield Torres L. F. \& Waga I., 1996, MNRAS, 279, 712\newline
Burud I., et al., 2000, preprint astro-ph/0007136\newline
Caldwell R. R., Dave R., \& Steinhardt P. J.,1998, Phys. Rev. Lett. 80, 1582%
\newline
Courbin F., Lidman P. \& Magain P., 1998, A\&A, 330, 57\newline
de Bernardis P., et al., 2000, Nature, 404, 995\newline
Demianski M., de Ritis R., Marino A. A., Piedipalumbo E., 2000, preprint
astro-ph/0004376\newline
Dyer C. C. \& Roeder R. C., 1972, ApJ, 174, L115\newline
Dyer C. C. \& Roeder R. C., 1974, ApJ, 189, 167\newline
Fassnacht C. D., et al., 1999, ApJ, 527, 513, preprint astro-ph/9907257%
\newline
Ferreira P. G. \& Joyce M., 1988, Phys. Rev. D58, 2350\newline
Frieman J., Hill C. T., Stebbins A. \& Waga I., 1995, Phys. Rev. Lett. 75,
2077\newline
Impey C. D., et al., 1998, ApJ, 509, 551.\newline
Koopmans L. V. E. \& Fassnacht C. D., 1999, ApJ, 527, 513, preprint
astro-ph/9907258\newline
Koopmans L. V. E., de Bruyn A. G., Xanthopoulos E. \& Fassnacht C. D., 2000,
A\&A, 356, 391, preprint astro-ph/0001533\newline
Kundi\'{c} T., et al., 1997,ApJ, 482, 75\newline
Lange A. E., et al.,2000, astro-ph/0005004, Phys. Rev. D submitted\newline
Leh\'{a}r J., et al., 1999, preprint astro-ph/9909072\newline
Lovell J. E. J., et al., 1998, ApJ, 508, 51, preprint astro-ph/9809301%
\newline
Patnaik A. R., Porcas R. W. \& Browne I. W. A., 1995, MNRAS, 274, L5\newline
Perlmutter S., et al., 1999, Ap.J., 517, 565\newline
Ratra B. \& Peebles P. J. E., 1988, Phys. Rev., D37, 3406.\newline
Riess A. G., et al., 1998, Ap.J., 116, 1009\newline
Schechter P. L., et al., 1996 preprint astro-ph/9611051\newline
Schneider P., Ehlers J. \& Falco E. E., 1992, ''Gravitational Lenses''
Springer-Verlag New York\newline
Silveira V., \& Waga I., 1997, Phys. Rev., D56, 4625\newline
Vilenkin A., 1984, Phys. Rev. Lett., 53, 1016\newline
Waga I. \& Miceli A. PaM. R., 1999, Phys. Rev., D 59, 1035, preprint
astro-ph/9811460\newline
Wetterich C., 1995, A\& A, 301, 321\newline
Wiklind T. \& Combes F., 1999, preprint astro-ph/9909314\newline
Wisotzki L., Wucknitz O., Lopez S. \&\ S\o rensen A. N., 1998
astro-ph/9810181 preprint\newline
Witt H. J., Mao S. \& Keeton C., R., 2000, preprint astro-ph/0004069\newline

\bigskip

%\section{\protect\bigskip Figure Caption}
\newpage

\begin{figure}[tbp]\epsfysize 8in
\epsfbox{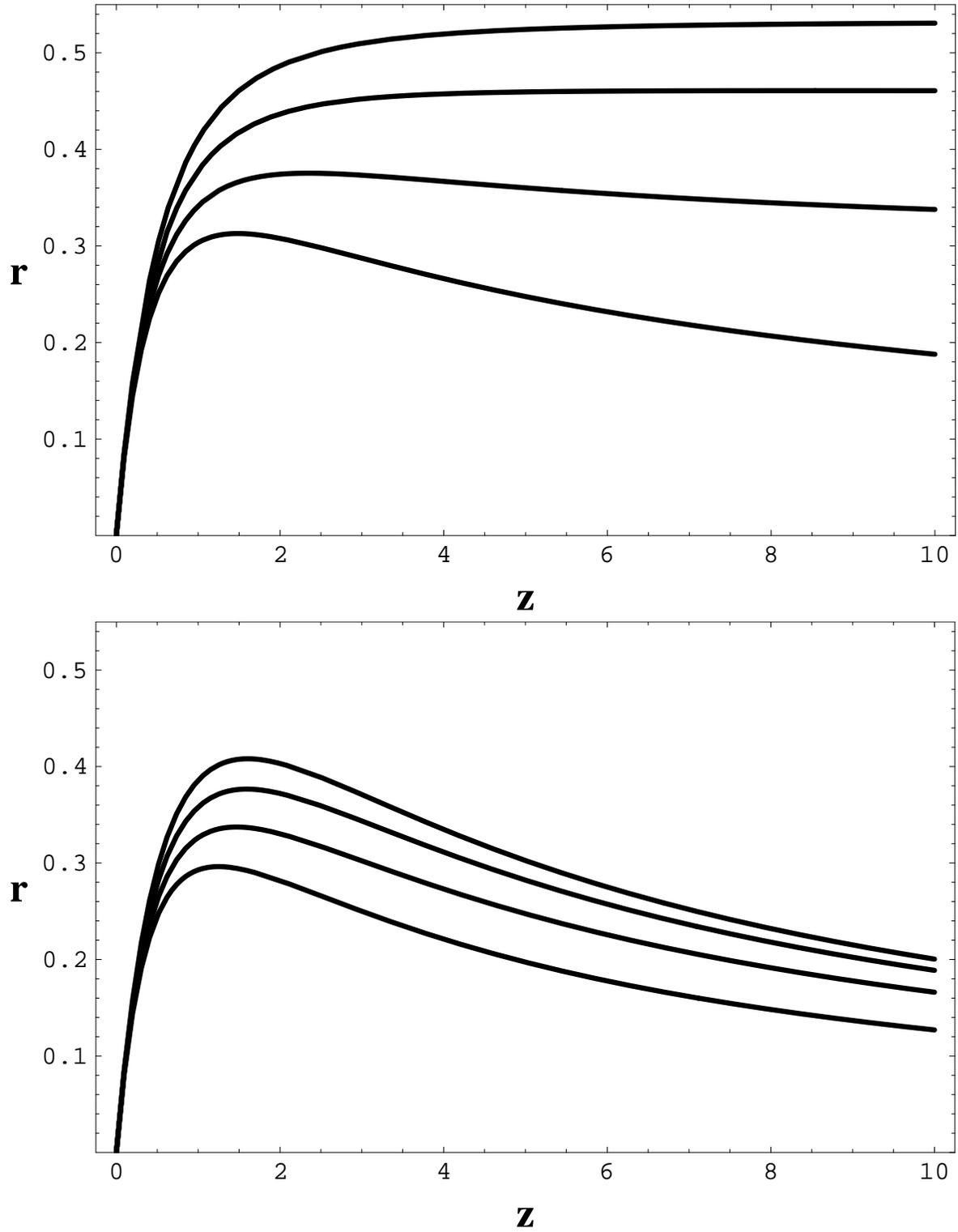}
\caption{ Top panel: DR distance for an empty beam ($\protect\alpha =0$)
with $\Omega _{m}=0.3$ and, from top to bottom, $m=0,1,2,3$. Bottom panel:
the same for a filled beam ($\protect\alpha =1$). Notice that for $\protect%
\alpha =0$ there is quite a larger dispersion among the distances. }
\end{figure}

\newpage

%\bigskip

\begin{figure}[tbp]\epsfysize 4in
\epsfbox{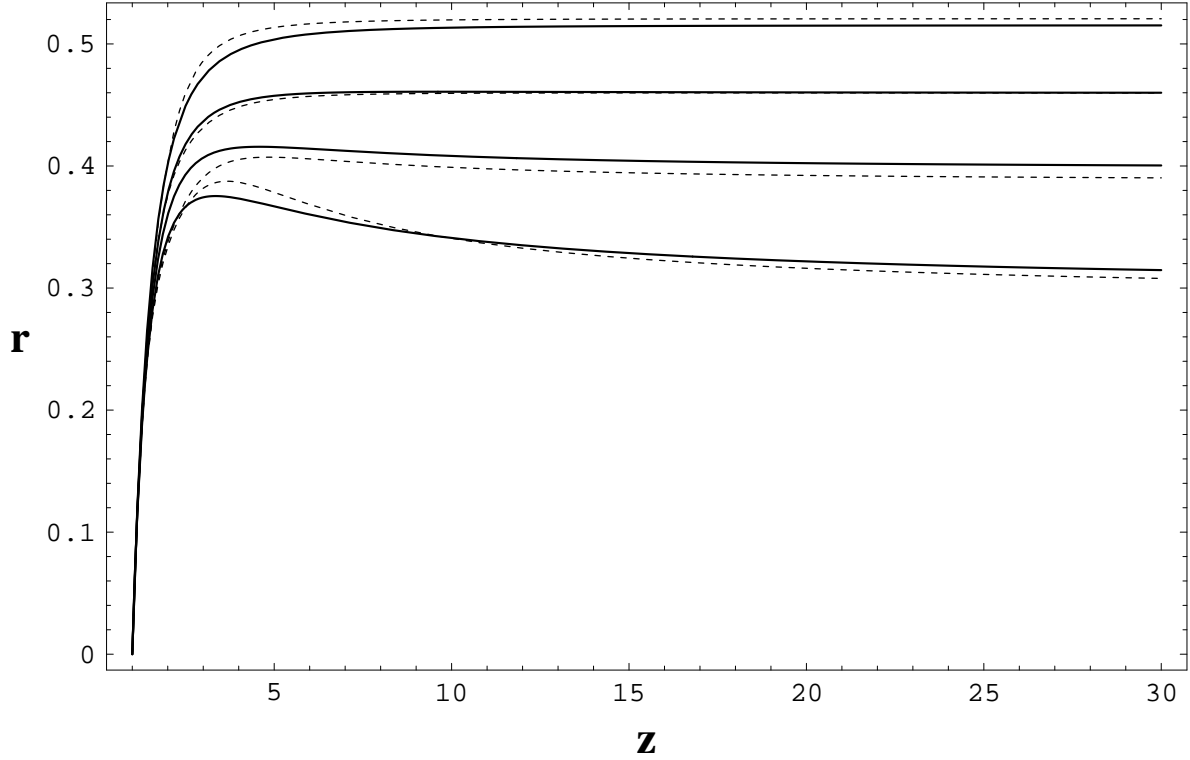}
\caption{ The solid lines are the numerical solutions for $\Omega _{m}=0.3$
and, from top to bottom, for $m=0.3;1;1.5;2.$ The dashed lines are our fits.}
\end{figure}
\newpage

%\bigskip

\begin{figure}[tbp]\epsfysize 8in
\epsfbox{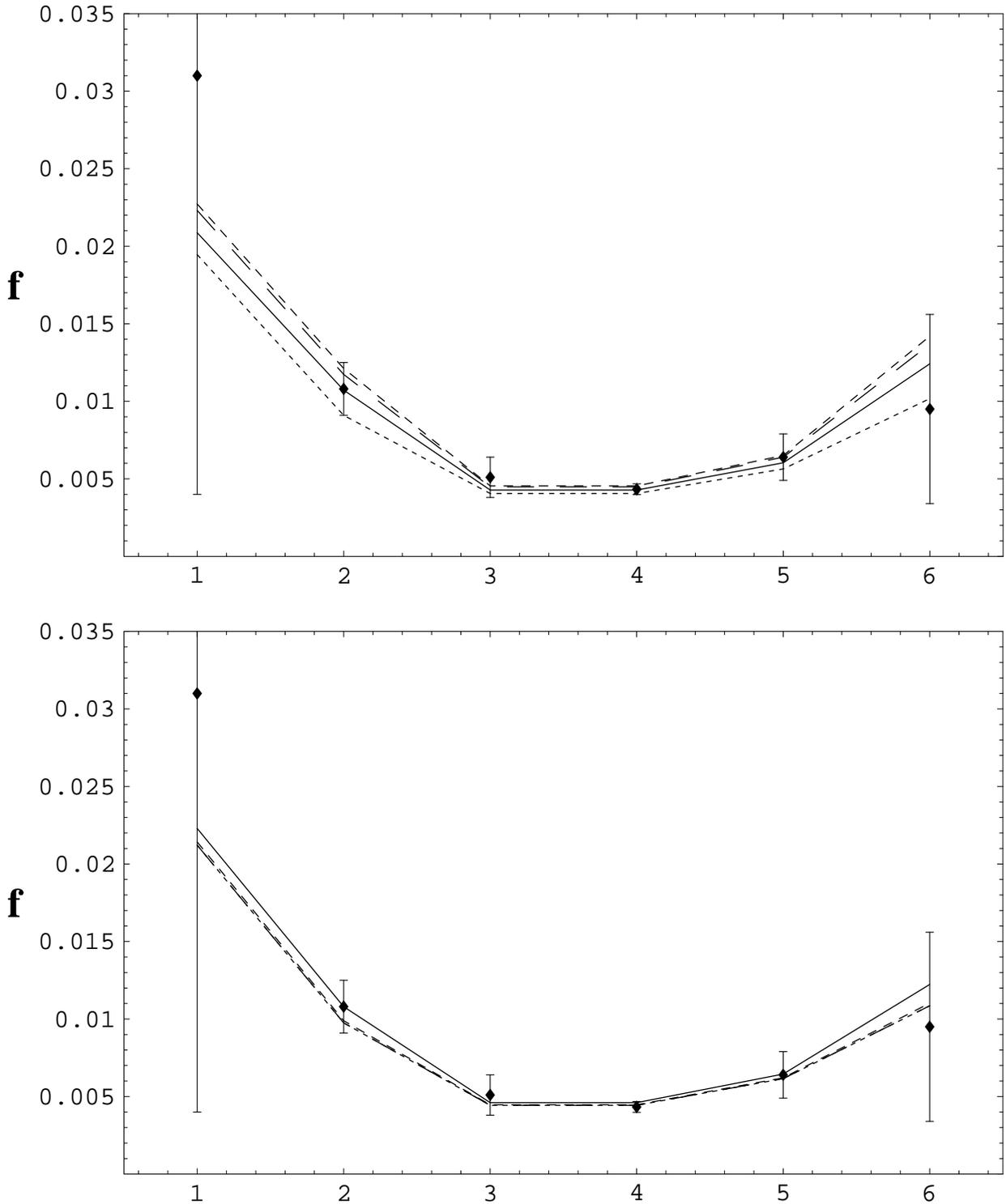}
\caption{ Top panel: comparison of the measured time delays $f$ of Table II
(points with error bars) with the theoretical time delays (lines) for an
empty beam and $H_{0}=71$\ km/s/Mpc. The theoretical parameters are: $\Omega
_{m}=0.1$ \ and $m=0$ (solid line) and $m=3$\ (dotted line); $\Omega
_{m}=0.9 $ and $m=0$ (short-dashed line) and $m=3$\ (long-dashed line).
Bottom panel: the same as above but for a filled beam and $H_{0}=64$%
{\protect\small \ }km/s/Mpc.}
\end{figure}
\newpage

%\bigskip

\begin{figure}[tbp]\epsfysize 8in
\epsfbox{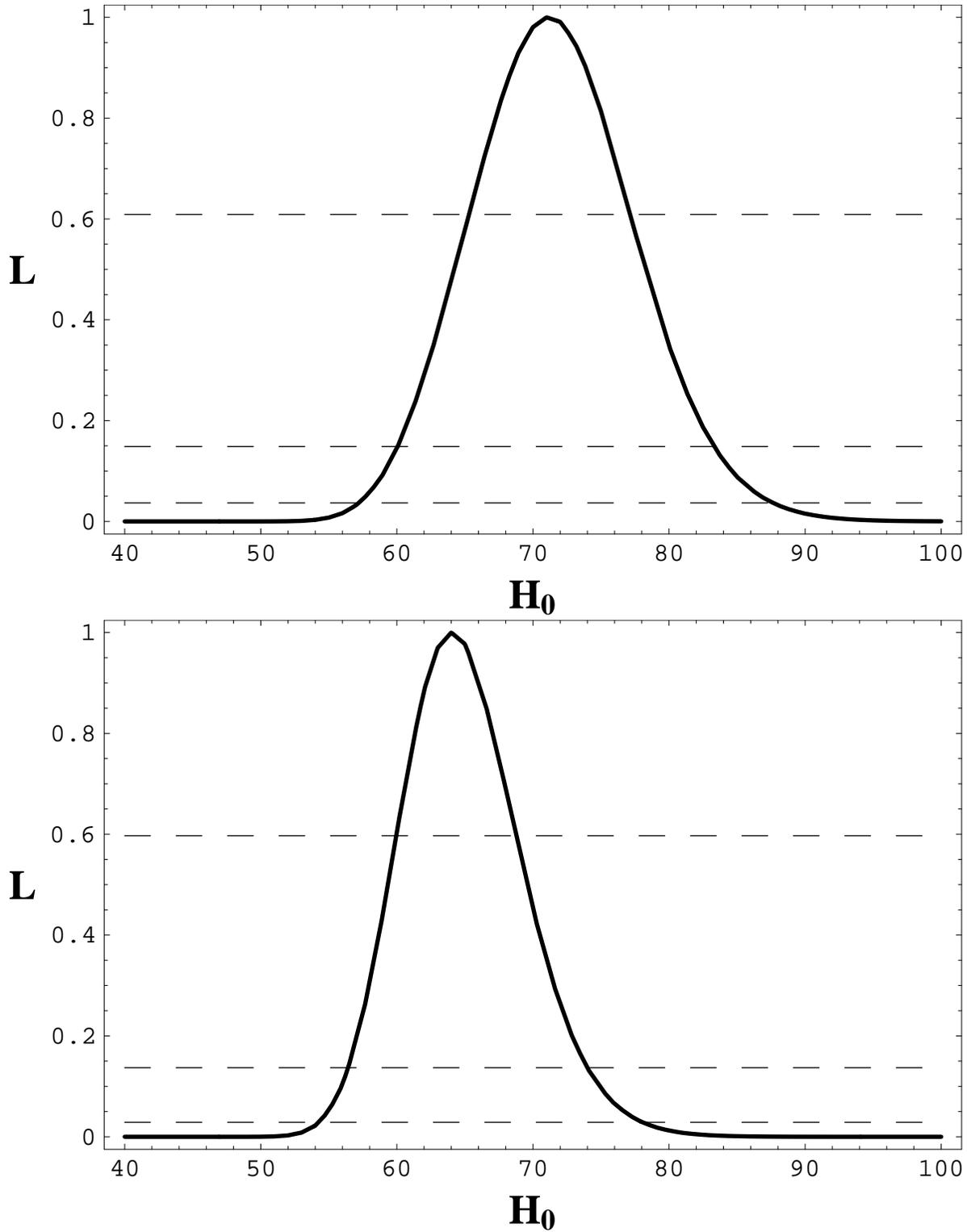}
\caption{ Top panel: The marginalized likelihood $L(H_{0})$ for $\protect%
\alpha =0.$ The dashed lines mark the 1$\protect\sigma ,$2$\protect\sigma $
and 3$\protect\sigma $ levels. Bottom panel: Same as above but for $\protect%
\alpha =1.$}
\end{figure}
\newpage

%\bigskip

\begin{figure}[tbp]\epsfysize 8in
\epsfbox{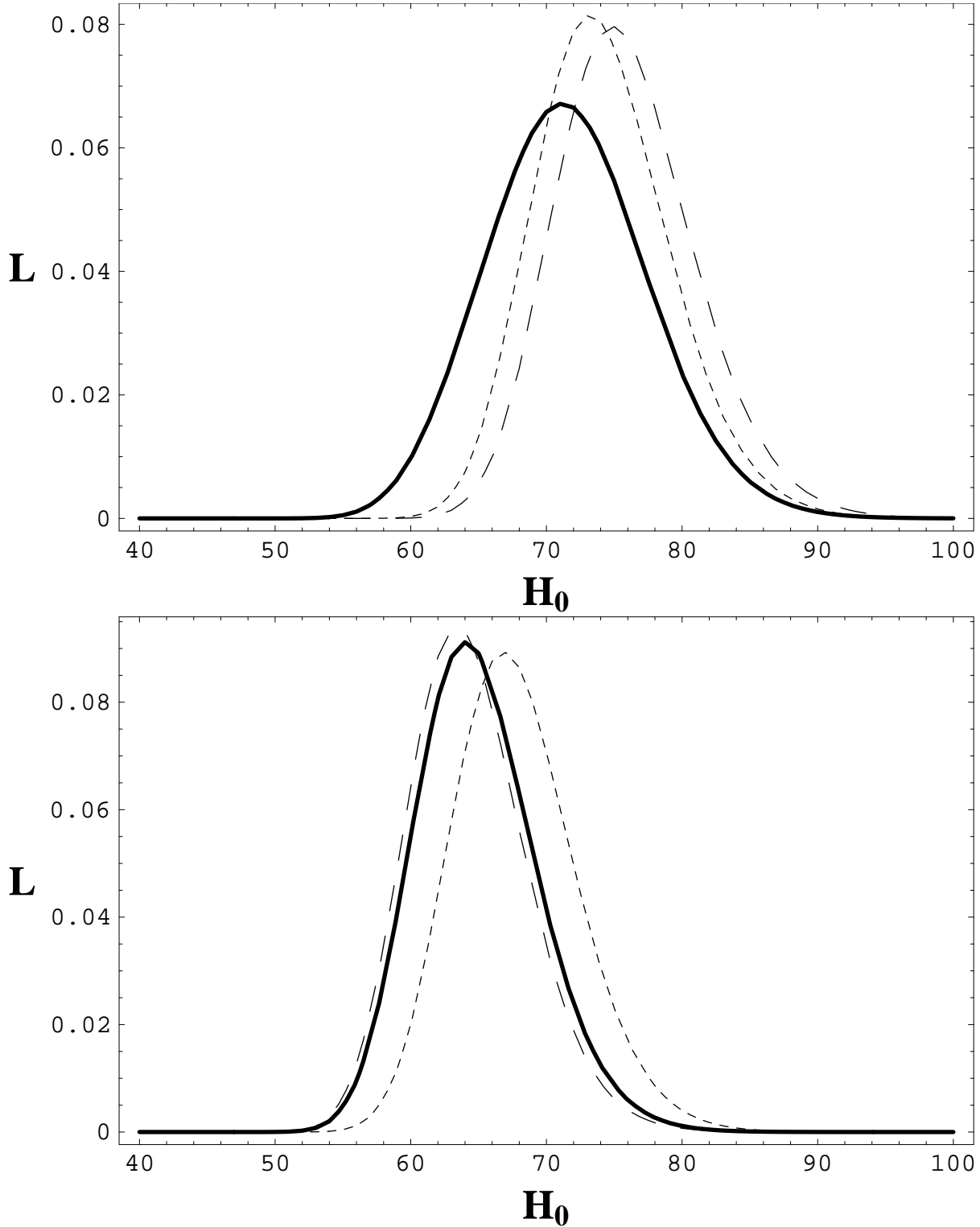}
\caption{ Top panel: comparison of the fully marginalized likelihood $%
L(H_{0})$ (solid line) for $\protect\alpha =0$ with the likelihood obtained
fixing the cosmology to two standard cases: pure cosmological constant ( $%
\Omega _{m}=0.3$ , $m=0,$ short-dashed line) and no quintessence ($\Omega
_{m}=1$ , $m=3,$ long-dashed line). Notice the significant shift. Bottom
panel: same as above but for $\protect\alpha =1.$ Here the shift is reduced.}
\end{figure}
\newpage

%\bigskip

\begin{figure}[tbp]\epsfysize 8in
\epsfbox{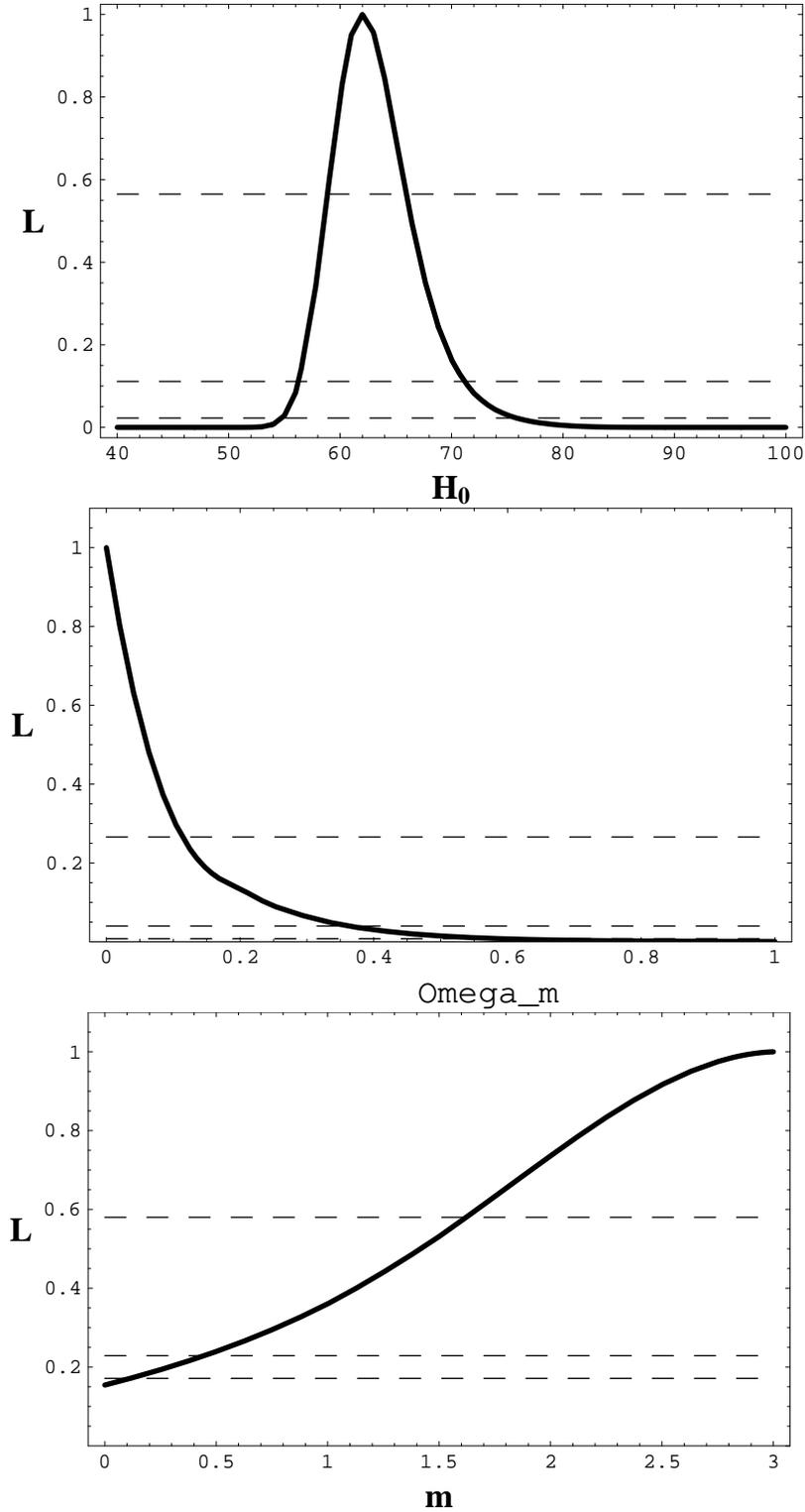}
\caption{ Simulation with a 10\% error on time delays assuming $\protect%
\alpha =0$. Top panel: marginalized likelihood of $H_{0}$ $.$ The dashed
lines are the levels at 1$\protect\sigma $,2$\protect\sigma $ and 3$\protect%
\sigma .$ Central panel: marginalized likelihood of $\Omega _{m}$ $.$ Bottom
panel: marginalized likelihood of $m$ $.$}
\end{figure}

\end{document}